\documentclass[aps,amsmath,amssymb,nofootinbib,twocolumn,showpacs,
               superscriptaddress]{revtex4}
\usepackage{graphicx}
\usepackage{amssymb}
\newcommand{\mb}[1]{\mbox{\scriptsize #1}}

\begin{document}

\title{Gluonic phase versus LOFF phase in two-flavor quark matter}

\author{O. Kiriyama}
\email{kiriyama@th.physik.uni-frankfurt.de}
\affiliation{Institut f\"ur Theoretische Physik, 
J.W.\ Goethe-Universit\"at, D-60438 Frankfurt am Main, Germany}

\author{D. H. Rischke}
\email{drischke@th.physik.uni-frankfurt.de}
\affiliation{Institut f\"ur Theoretische Physik, 
J.W.\ Goethe-Universit\"at, D-60438 Frankfurt am Main, Germany}
\affiliation{
Frankfurt Institute for Advanced Studies, 
J.W.\ Goethe-Universit\"at, D-60438 Frankfurt am Main, Germany}

\author{I. A. Shovkovy}
\email{shovkovy@th.physik.uni-frankfurt.de}
  \altaffiliation[on leave from ]{%
       Bogolyubov Institute for Theoretical Physics,
       03143, Kiev, Ukraine}
\affiliation{
Frankfurt Institute for Advanced Studies, 
J.W.\ Goethe-Universit\"at, D-60438 Frankfurt am Main, Germany}
\affiliation{Department of Physics, Western Illinois University, Macomb, IL 61455, USA}%

\begin{abstract}
We study the gluonic phase in a two-flavor color superconductor  
as a function of the ratio of the gap over the chemical potential mismatch,
$\Delta/\delta\mu$. We find that the gluonic phase resolves
the chromomagnetic instability encountered in a
two-flavor color superconductor for 
$\Delta/\delta \mu < \sqrt{2}$. 
We also calculate approximately
the free energies of the gluonic phase and the single plane-wave LOFF
phase and show that the former is favored over the latter for 
a wide range of coupling strengths.
\end{abstract}

\date{\today}
\maketitle

It is widely accepted that sufficiently cold and dense quark matter 
is a color superconductor \cite{CSC}. The most likely and, probably, 
the only place where color superconductivity can exist in the universe 
is the interior of compact stars. Thus, studies of phases of quark 
matter under conditions realized in the bulk of compact stars (i.e., 
color and electric charge neutrality, and $\beta$-equilibrium) have 
recently attracted a great deal of interest. The density regime of 
relevance for compact stars is up to a few times the normal 
nuclear density $\rho_0 \simeq 0.16~{\rm fm}^{-3}$. In this ``moderate'' 
density regime, the studies of QCD-motivated effective theories are 
most useful and have revealed a rich phase structure
\cite{phased1,phased2,phased3}. 

One of the most striking features 
of neutral and $\beta$-equilibrated color-superconducting phases 
is unconventional cross-flavor Cooper pairing of quarks with 
the possibility of gapless superconductivity, e.g.,  in the form 
of the gapless 2SC (g2SC) phase \cite{Shovkovy2003} or the gapless 
color-flavor-locked (gCFL) phase \cite{Alford2003}. It was, however, quickly 
realized that the 2SC/g2SC phases suffer from a 
chromomagnetic instability, indicated by imaginary Meissner screening 
masses of some gluons \cite{Huang2004}. 
In the 2SC phase, these instabilities occur when
the ratio of the gap over the mismatch of the chemical potential,
$\Delta/\delta \mu$, decreases below a value $\sqrt{2}$.
Similar instabilities were found also in the gCFL phase 
\cite{Casalbuoni2004}. 

Resolving the chromomagnetic instability and clarifying 
the nature of the true ground state of dense quark matter are the most 
pressing tasks in the study of color superconductors. It was proposed 
that the chromomagnetic instability in two-flavor quark matter 
can be removed by the formation of a single plane-wave LOFF state 
\cite{Giannakis2004,Giannakis2005,Huang2005,Hong2005} (first studied 
by Larkin and Ovchinnikov \cite{LO}, and Fulde and Ferrell \cite{FF} 
in the context of solid state physics, 
and by Alford, Bowers, and Rajagopal \cite{ABR} 
for cold, dense quark matter), or a gluonic phase with 
vector condensation in the ground state \cite{Gorbar2005a}. (For a 
recent discussion of this issue, see also Refs.~\cite{Iida2006,Fukush2006}.) 
Alternatives include a mixed phase \cite{RedRup} and, in the 
case of three-flavor quark matter, also phases with spontaneously 
induced meson supercurrents \cite{supercurrent}. While the neutral 
LOFF state is free from the chromomagnetic instability in the 
weak-coupling regime \cite{Giannakis2005}, this is, in fact, not the case 
for somewhat larger values of the coupling \cite{Gorbar2005b}. 
At the same time, the gluonic phase can resolve the instability there. 
So far, however, the gluonic phase phase has been studied only around 
the critical point $\Delta/\delta \mu = \sqrt{2}$ \cite{Gorbar2005a}. 

The gluonic phase and the LOFF phases are currently viewed as the 
most likely candidates for resolving the chromomagnetic instability 
and, thus, for the true ground state of two-flavor color-superconducting 
quark matter. (Here we exclude the possibility of phase separation 
\cite{RedRup} which may have limitations of its own and deserves a separate
in-depth study.) In order to see which of the two proposed phases is 
actually preferred, one first has to extend the analysis of 
Ref.~\cite{Gorbar2005a} to a computation of the free energy 
away from the critical point $\Delta/\delta \mu = \sqrt{2}$, 
and then compare the results to the free energy of the single-plane 
LOFF state.\footnote{Note that while the comparison with a multi-plane wave 
LOFF states would be more desirable, the corresponding free energy 
cannot be easily estimated within a microscopic approach. 
For current state-of-the-art calculations using an effective theory see 
Ref.~\cite{Rajagopal2006}.} This is done in the present work. We 
qualitatively confirm the results of Ref.~\cite{Fukush2006} 
and extend them by (approximately) including the neutrality condition 
and explicitly comparing the free energy of the gluonic phase to that of 
the single plane-wave LOFF phase as a function of the coupling strength.

In order to study various phases of two-flavor quark matter, we use a gauged 
Nambu--Jona-Lasinio (NJL) model with massless up and down quarks:
\begin{eqnarray}
{\cal L}&=&\bar{\psi}(iD\hspace{-7pt}/+\hat{\mu}\gamma^0)\psi
+G_D\left(\bar{\psi}i\gamma_5\varepsilon\epsilon^bC\bar{\psi}^T\right)
\left(\psi^T Ci\gamma_5\varepsilon\epsilon^b\psi\right)\nonumber\\
&&-\frac{1}{4}F_{\mu\nu}^{a}F^{a\mu\nu},
\end{eqnarray}
where the quark field $\psi$ carries flavor ($i,j=1,\ldots N_f$ 
with $N_f=2$) and color ($\alpha,\beta=1,\ldots N_c$ with $N_c=3$) 
indices, $C$ is the charge conjugation matrix; 
$(\varepsilon)^{ik}=\varepsilon^{ik}$ and 
$(\epsilon^b)^{\alpha\beta}=\epsilon^{b\alpha\beta}$ 
are the antisymmetric tensors in flavor and color spaces, 
respectively. The covariant derivative and the field 
strength tensor are defined as
\begin{subequations}
\begin{eqnarray}
D_{\mu} &=& \partial_{\mu}-igA_{\mu}^{a}T^{a},\\
F_{\mu\nu}^{a} &=& \partial_{\mu}A_{\nu}^{a}-\partial_{\nu}A_{\mu}^{a}
+gf^{abc}A_{\mu}^{b}A_{\nu}^{c}.
\end{eqnarray}
\end{subequations}
To evaluate loop diagrams we use a three-momentum cutoff 
$\Lambda$. Hence, the model has two phenomenological model parameters, 
the cutoff $\Lambda$ and the diquark coupling $G_D$. We use 
$\Lambda=653.3$~MeV throughout this paper, but we consider $G_D$ as a 
free parameter. Henceforth, in order to specify the diquark coupling we use 
$\Delta_0$ which is the value of the 2SC gap at $\delta\mu=0$ 
(see below).

In $\beta$-equilibrated neutral 2SC/g2SC matter, the elements of the 
diagonal matrix of quark chemical potentials $\hat{\mu}$ are given 
by
\begin{subequations}
\begin{eqnarray}
\mu_{ur}&=&\mu_{ug}=\bar{\mu}-\delta\mu,\\
\mu_{dr}&=&\mu_{dg}=\bar{\mu}+\delta\mu,\\
\mu_{ub}&=&\bar{\mu}-\delta\mu-\mu_8,\\
\mu_{db}&=&\bar{\mu}+\delta\mu-\mu_8,
\end{eqnarray}
\end{subequations}
with
\begin{eqnarray}
\bar{\mu}=\mu-\frac{\delta\mu}{3}+\frac{\mu_8}{3}~,\qquad
\delta\mu=\frac{\mu_e}{2}.
\end{eqnarray}
In a gauge theory, the self-consistent solution 
of the Yang-Mills equations requires 
background gauge fields \cite{Gerhold2003}. 
These can be viewed as electric- and color-chemical potentials 
which ensure electric and color-charge neutrality of the system. 
Note that a generalization of this holds true even in the case of 
inhomogeneous phases. Then, of course, the corresponding 
fields would not be constant in space. Instead, they would have a constant 
central value contribution and, on top of it, a coordinate-dependent 
modulation describing color-electric fields induced by the 
inhomogeneities.\footnote{In the special case of a mixed phase, e.g.,
a color-electric field is generated around the boundary layer between 
the two phases and prevents the generation of a color-electric 
current across this layer.} The constant contribution would take 
care of the {\it global} neutrality, while the modulation describes the 
local field needed to prevent the {\it local} flow of currents.

On the other hand, in NJL-type models without dynamic gauge fields, 
one has to ensure electric and color-charge neutrality 
by introducing appropriate chemical potentials by hand \cite{BubSho}. 
In the case of the 2SC/g2SC phases, we only require an electron chemical 
potential $\mu_e$, and a color-chemical potential $\mu_8$ which ensures 
that the color-charge density $n_8$ is zero. In principle, in other 
phases like the gluonic phase one has to check that no other 
color-charge density is non-vanishing and necessitates the introduction 
of a respective color-chemical potential. 
Indeed, the gluonic phase introduced in Ref. \cite{Gorbar2005a} requires a 
non-vanishing temporal component of the gluon field of the third 
color, $\langle A_0^3 \rangle$. In our gauged NJL model, this is 
equivalent to a non-vanishing color-chemical potential $\mu_3$ besides 
$\mu_8$. 
In this first exploratory study, however, we use the fact that 
both $\mu_3$ and $\mu_8$ are known to be numerically small 
and we simply neglect them.

In Nambu-Gor'kov space, the inverse full quark propagator 
$S^{-1}(p)$ is written as
\begin{eqnarray}
S^{-1}(p)=\left(
\begin{array}{cc}
(S_0^+)^{-1} & \Phi^- \\
\Phi^+ & (S_0^-)^{-1} 
\end{array}
\right),
\end{eqnarray}
with
\begin{subequations}
\begin{eqnarray}
&&(S_0^+)^{-1}=\gamma^{\mu}p_{\mu}+(\bar{\mu}-\delta\mu\tau^3)\gamma^0
+g\gamma^{\mu}A_{\mu}^{a}T^{a},\\
&&(S_0^-)^{-1}=\gamma^{\mu}p_{\mu}-(\bar{\mu}-\delta\mu\tau^3)\gamma^0
-g\gamma^{\mu}A_{\mu}^{a}T^{aT},
\end{eqnarray}
\end{subequations}
and
\begin{eqnarray}
\Phi^- = -i\varepsilon\epsilon^b\gamma_5\Delta~,\qquad
\Phi^+ = -i\varepsilon\epsilon^b\gamma_5\Delta.
\end{eqnarray}
Here $\tau^3=\mbox{diag}(1,-1)$ is a matrix in flavor space. Following the 
usual convention, we choose the diquark condensate to point 
in the third (blue) direction in color space. 

In the one-loop approximation, the free energy 
of two-flavor quark matter at $T=0$ is given by
\begin{eqnarray}
V_g=\frac{\Delta^2}{4G_D}
-\frac{1}{2}\int\frac{d^4p}{(2\pi)^4i}\ln\mbox{Det}S^{-1}(p),
\label{eqn:omega1}
\end{eqnarray}
where ``Det'' stands for the determinant in Dirac, flavor, color, and 
Nambu-Gor'kov space. Unlike the free energy in Ref.~\cite{Gorbar2005a}, 
Eq.~(\ref{eqn:omega1}) does not have a quartic term in $A_{\mu}^{a}$.
This is because we neglected the color-chemical potential $\mu_8$ 
and, in addition, we take into account only one dynamic gluonic field 
(see below).

In the gluonic phase \cite{Gorbar2005a}, the chromomagnetic 
instability at $\Delta/\delta\mu < \sqrt{2}$ triggers a 
non-vanishing vacuum expectation value of the spatial 
component of 
\begin{eqnarray}
K_{\mu} = \frac{1}{\sqrt{2}}
\left(
\begin{array}{c}
A_{\mu}^{4}-iA_{\mu}^{5}\\
A_{\mu}^{6}-iA_{\mu}^{7} 
\end{array}
\right).
\end{eqnarray}
(For a simpler version of such a phenomenon, see also 
Ref.~\cite{Gusynin2003}.) Using the $\mbox{SO}(3)_{\mb{rot}}$ 
rotational symmetry and the $\mbox{SU}(2)_c$ color symmetry, one 
can choose $B \equiv g\langle A_z^{6} \rangle \neq 0$ without loss 
of generality. Consequently, the non-zero vacuum expectation 
value of $B$ breaks $\mbox{SO}(3)_{\mb{rot}}$, leaving 
only $\mbox{SO}(2)_{\mb{rot}}$ \cite{Gorbar2005a}. Furthermore, 
a non-vanishing $B$ together with $\Delta$ and $\mu_e$ breaks 
the original symmetry of QCD down to 
\begin{eqnarray}
\mbox{U}(1)_{\tilde{Q}}\otimes\mbox{U}(1)_{\tau_L^3}
\otimes\mbox{U}(1)_{\tau_R^3}\otimes\mbox{SO}(2)_{\mb{rot}}\, ,
\end{eqnarray}
where $\mbox{U}(1)_{\tau_{L/R}^3}$ is a subgroup of the 
$\mbox{SU}(2)_{L/R}$ chiral symmetry and the charge $\tilde{Q}$ is given by 
\begin{eqnarray}
\tilde{Q}=Q_f \otimes \openone_c
-\openone_f \otimes T^{3}
-\frac{1}{\sqrt{3}}\openone_f\otimes T^{8},
\end{eqnarray}
with $Q_f=\mbox{diag}(\frac23,-\frac13)$ being the flavor matrix of 
the electric charges of quarks. 

The reduced symmetry of the ground state with $B\neq 0$ allows 
for additional condensates, $C=g\langle A_z^{1} \rangle \neq 0$ 
and $D=g\langle A_0^{3} \rangle \neq 0$. In fact, as discussed in 
Ref.~\cite{Gorbar2005a}, such condensates are required by the equation 
of motion. 
As discussed above, the gluonic field $D$ is nothing but a color chemical 
potential $\mu_3$. The field $C$, on the other hand, induces electric 
superconductivity in the ground state and, therefore, is physically 
more interesting. However, including all three gluonic fields makes 
the analysis quite involved. In this work, we retain only the $B$ field 
that is directly connected to the Meissner masses of gluons 4--7 
and, thus, is the most relevant field for the chromomagnetic instability. 

It is straightforward to show that the mass of the $B$ field at 
$B=0$ (i.e., in the 2SC/g2SC phases) coincides with the Meissner 
screening masses of gluons of adjoint color 4--7 calculated 
in the hard-dense-loop (HDL) approximation \cite{Huang2004},
\begin{eqnarray}
M_B^2 &=& \frac{\partial^2 V_g}{\partial B^2}\bigg{|}_{B=0}
=\frac{\bar{\mu}^2}{6\pi^2}
\Bigg[1-\frac{2\delta\mu^2}{\Delta^2} \nonumber\\
&&+2\frac{\delta\mu\sqrt{\delta\mu^2-\Delta^2}}{\Delta^2}
\theta(\delta\mu-\Delta)\Bigg].\label{eqn:mm}
\end{eqnarray}
In order to derive this expression we neglected 
terms of order $O(\bar{\mu}^2/\Lambda^2)$ and $O(\Delta^2/\mu^2)$. 

\begin{figure}
\includegraphics[width=0.48\textwidth]{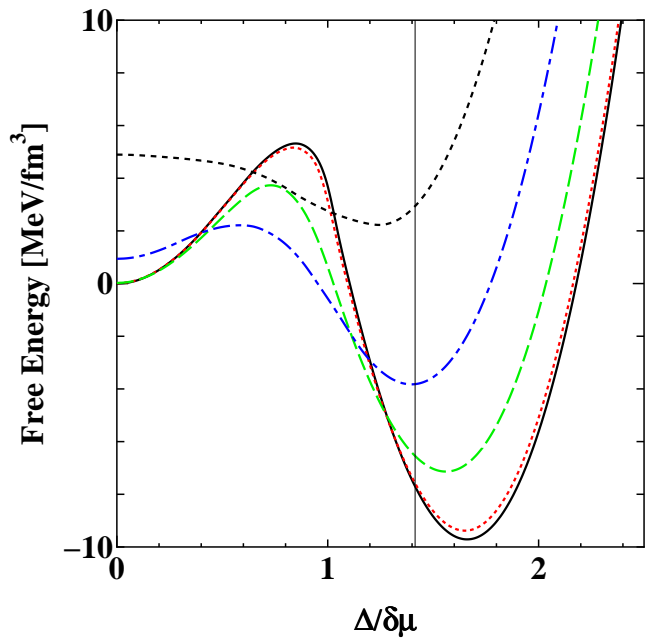}
\caption{The free energy (\ref{eqn:omega2}) measured with respect 
to the normal phase at $B=0$, as a function of 
$\Delta/\delta\mu$ for $B=0$~MeV (solid), 50~MeV (dotted), 
150~MeV (dashed), 250~MeV (dot-dashed), 350~MeV (short-dashed).
The thin vertical line denotes the critical point of the 
chromomagnetic instability $\Delta/\delta\mu=\sqrt{2}$. We 
used $\bar{\mu}=500~{\rm MeV}$ and $\delta\mu=80~{\rm MeV}$,
and the value of the coupling constant has been chosen so 
that $\Delta_0=132$~MeV. The contribution of electrons is 
not included.}
\label{Figure1}
\end{figure}
\begin{figure}
\includegraphics[width=0.48\textwidth]{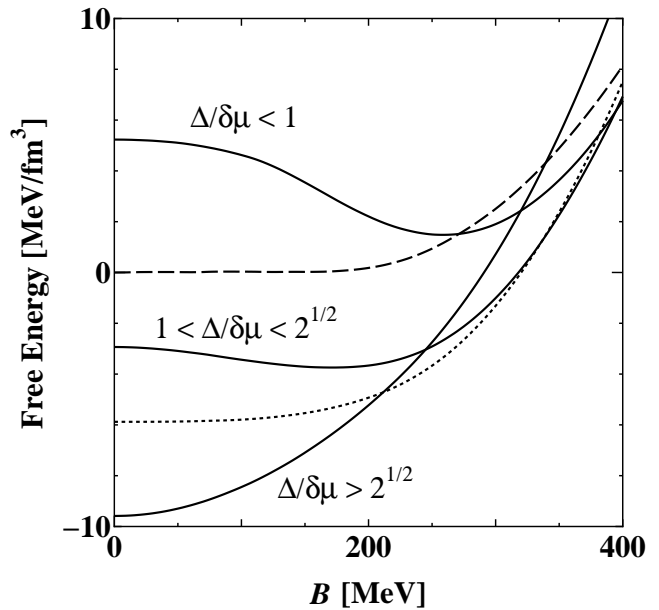}
\caption{The free energy (\ref{eqn:omega2}) measured with respect 
to the normal phase at $B=0$, as a function of $B$ 
in three regimes with different values of $\Delta/\delta\mu$. 
The solid lines correspond to the cases of $\Delta/\delta\mu=0.8,~1.2,~1.6$ 
(top to bottom). 
The dashed and dotted line denote the free energy 
at $\Delta/\delta\mu=0$ and at the critical point 
$\Delta/\delta\mu\simeq\sqrt{2}$, respectively. 
Other parameters are the same as in Fig.~\ref{Figure1}.}
\label{Figure2}
\end{figure}

In order to study the effect of the condensate field $B$ on the free 
energy of the 2SC/g2SC phases, we calculate the difference of the
thermodynamic potentials in a dense medium and in vacuum at the same 
value of $B$,
\begin{eqnarray}
\Omega_g\equiv
V_g(\Delta,B,\delta\mu,\mu)
-V_g(0,B,0,0).\label{eqn:omega2}
\end{eqnarray}
In a gauge theory, this subtraction in the one-loop free energy 
also takes care of the renormalization of the gauge coupling constant. 
As a result, the cutoff dependence of the free energy can be
completely removed in this approximation.

Let us look at the free energy $\Omega_g$ in detail. 
Fig.~\ref{Figure1} shows the free energy $\Omega_g$ 
(measured with respect to the normal phase at $B=0$) as a 
function of $\Delta/\delta\mu$. The results are plotted for 
$\bar{\mu}=500$~MeV and 
$\delta\mu=80$~MeV, with the diquark coupling chosen so that 
$\Delta_0=132$~MeV. Here we do not restrict $\Delta/\delta\mu$ 
to its physical value, determined by the stationary point of 
$\Omega_g(\Delta)$, but treat it as a free parameter. 

Several important features of the free energy as a function 
of the $B$ field are evident from Fig.~\ref{Figure1}. When 
$\Delta/\delta\mu > \sqrt{2}$, one can see that the free energy 
monotonically increases with $B$. 
(Strictly speaking, 
since our model reproduces the HDL result 
only up to terms of order 
$O(\bar{\mu}^2/\Lambda^2)$ and $O(\Delta^2/\mu^2)$, 
the actual critical point is somewhat lower than $\sqrt{2}$.) 
In other words, $\Omega_g(B)$ 
has a global minimum at $B=0$, and the 2SC phase is stable against 
gluon condensation in this regime. This is also clear from a
different representation of the results, shown in Fig.~\ref{Figure2}.

When $\Delta/\delta\mu < \sqrt{2}$, on the other hand,
we observe the onset of the chromomagnetic instability. 
For small $B$, the free energy first decreases with increasing $B$ 
and then grows at larger $B$. This can be 
seen clearly in Fig.~\ref{Figure2}. The behavior of 
$\Omega_g$ agrees well with Eq.~(\ref{eqn:mm}) at small 
$B$. In this regime, the 2SC/g2SC phase is no longer the 
ground state. It is unstable with respect to the formation 
of a non-zero $B$ condensate, i.e., the so-called 
gluonic phase. The corresponding ground state is 
determined by the minimum of $\Omega_g(B)$. The Meissner 
masses squared, which are given by the curvature of the free 
energy at the minimum, are non-negative in this state. 
[It is interesting to note that, although the free energy 
in the normal phase, $\Delta/\delta\mu=0$, 
cf. dashed line in Fig.~\ref{Figure2}, 
increases with $B$, this increase is extremely slow over quite 
a large range of $B$ values. The reason is that the quadratic term in 
the Taylor expansion of the free energy is vanishing, see Eq.~(\ref{eqn:mm}).]
Let us note that the results of Fig.~\ref{Figure2} 
are in qualitative agreement 
with those shown in Fig.~2 of Ref.~\cite{Fukush2006}.

{From} the results for the free energy it is clear that the gluonic 
phase resolves the chromomagnetic instability of the 2SC/g2SC phases. 
A neutral LOFF state is another candidate for the solution to the 
instability: it has been shown that such a state is free from the 
chromomagnetic instability, however, only in the weak-coupling regime 
\cite{Giannakis2005} (for strong coupling, 
this is not the case \cite{Gorbar2005b}). 
In order to determine the energetically most favored state, 
it is necessary to compare the 
free energies of the 2SC/g2SC phases, the neutral LOFF state and 
the gluonic phase. 

To this end we use the following approximation derived 
in Ref.~\cite{Gorbar2005b} for the 2SC/g2SC phases 
and the neutral LOFF state:
\begin{eqnarray}
\Omega=\Omega_{\mb{2SC}}+\Omega_{\mb{g2SC/LOFF}},
\end{eqnarray}
where the 2SC and g2SC/LOFF parts of the free energy are given by
\begin{widetext}
\begin{subequations}
\label{eq:appomega}
\begin{eqnarray}
&&\Omega_{\mb{2SC}}=
-\frac{\mu_e^4}{12\pi^2}
-\frac{\mu_{ub}^4}{12\pi^2}
-\frac{\mu_{db}^4}{12\pi^2}
-\frac{\bar{\mu}^4}{3\pi^2}
+\frac{\Delta^2}{4G_D}
-\frac{\bar{\mu}^2\Delta^2}{\pi^2}
\ln\frac{4(\Lambda^2-\bar{\mu}^2)}{\Delta^2}
-\frac{\Delta^2}{\pi^2}\left(\Lambda^2-2\bar{\mu}^2\right),\\
&&\Omega_{\mb{g2SC/LOFF}}=\frac{2\bar{\mu}^2q^2}{\pi^2}
+\frac{\bar{\mu}^2}{\pi^2}
\left\{\frac{(q+\delta\mu)^3}{q}
\left[\frac{1}{2}(1-x_1^2)\ln\frac{1+x_1}{1-x_1}-x_1+\frac{2}{3}x_1^3\right]
+(q \to -q)\right\},
\end{eqnarray}
\end{subequations}
\end{widetext}
with the dimensionless parameter $x_1$ being
\begin{eqnarray}
x_1=\theta\left(1-\frac{\Delta^2}{(\delta\mu+q)^2}\right)
\sqrt{1-\frac{\Delta^2}{(\delta\mu+q)^2}}
\end{eqnarray}
and 
\begin{eqnarray}
q=|\vec{q}|~,~~
\vec{q}=\frac{g}{2\sqrt{3}}\langle\vec{A}^{8}\rangle \, .
\end{eqnarray}
Note that the wave vector of the diquark condensate $\vec{q}$ is 
equivalent to a gauge field condensate $\langle\vec{A}^{8}\rangle$ 
in the case of single plane-wave LOFF pairing. In 
Eq.~(\ref{eq:appomega}), the 2SC/g2SC part of the free energy is 
obtained by taking the $q \to 0$ limit. Also a non-zero color 
chemical potential $\mu_8$ has been neglected there. 

The free energy of a given phase can be computed by solving the 
gap equations, e.g., $\partial\Omega/\partial\Delta=0$ and 
$\partial\Omega/\partial q=0$, and the neutrality condition 
$\partial\Omega/\partial\delta\mu=0$. To simplify the calculations
in the gluonic phase, we evaluate the free energy approximately
as follows: (i) we obtain $\Delta^{\star}$ and $\delta\mu^{\star}$ 
in the 2SC/g2SC phase by solving the coupled set of equations 
$\partial\Omega/\partial\Delta=0$ and 
$\partial\Omega/\partial\delta\mu=0$, (ii) by using these solutions,
we calculate $\Omega_g(B,\Delta^{\star},\delta\mu^{\star})$ which is 
an approximate value for the free energy in the gluonic phase. 
For the densities of interests, 
$B$ is at most of the order of 100 MeV, 
whereas $q$ is of the order of tens of MeV. 
We performed a preliminary test of the quality of our approximation by 
varying $B$ from 0 to 300 MeV and computing the value of $\delta\mu$ 
necessary to ensure electric neutrality. We found that this value 
changes at most by 10\%.

\begin{figure}
\includegraphics[width=0.48\textwidth]{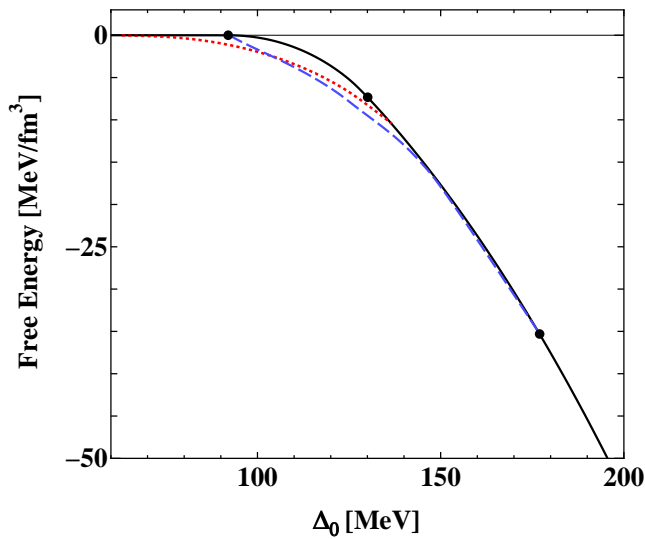}
\caption{The free energy of 
the neutral 2SC/g2SC phase (solid line), 
the neutral LOFF state (dotted line), 
and the gluonic phase (dashed line) 
as a function of $\Delta_0$. 
The three dots on the bold solid line 
($\Delta_0=92,~130,~177$~MeV from left to right) 
denote the edge of the g2SC window with the normal phase, 
the phase transition point between 2SC and g2SC phases ($\Delta=\delta\mu$), 
and the critical point of the chromomagnetic instability 
($\Delta=\sqrt{2}\delta\mu$). 
The quark chemical potential is taken to be $\mu=400$~MeV.}
\label{Figure3}
\end{figure}

We illustrate the comparison of the free energies of all three 
phases in Fig.~\ref{Figure3} (cf. Fig.~2 in Ref.~\cite{Gorbar2005b}).
We take $\mu=400$~MeV and choose the normal phase as a reference 
point for the free energy. In Ref.~\cite{Gorbar2005b}, it has been 
demonstrated that the neutral LOFF state is more stable than 
the 2SC/g2SC phases in the whole LOFF window 
$63~{\rm MeV}<\Delta_0<137~{\rm MeV}$, 
which includes the entire g2SC window 
$92~{\rm MeV}<\Delta_0<130~{\rm MeV}$. 
However, whereas the Meissner masses squared of gluons 4--7 
in the weakly coupled neutral LOFF state are positive 
\cite{Giannakis2004,Giannakis2005}, they remain negative in the 
intermediate and the strongly coupled regimes (at all values of 
$\Delta_0$ above 81~MeV) \cite{Gorbar2005b}. In contrast, the 
gluonic phase removes the instability and is energetically favored 
over the 2SC/g2SC phases in the whole window in which the instability 
takes place (at all values of $\Delta/\delta\mu$ below $\sqrt{2}$). 
The gluonic phase and the neutral LOFF state coexist in the region 
$92~{\rm MeV}<\Delta_0<137~{\rm MeV}$, but, as our results indicate, 
the gluonic phase is more stable than the neutral LOFF state in a 
wide region $\Delta_0 > 103$~MeV, which is close to the edge of the 
g2SC window with the normal phase. We argue therefore that the 
instability in the 2SC/g2SC phase is resolved by the formation 
of the gluonic phase.

In summary, we explored the gluonic phase away from the 
critical point $\Delta/\delta\mu=\sqrt{2}$. We demonstrated 
that the energetically favored state of a neutral two-flavor color 
superconductor is not the 2SC/g2SC phase in the intermediate- 
and strong-coupling regimes but the gluonic phase in which 
the dynamic gluonic field $B=g\langle A_z^{6} \rangle$ 
acquires a vacuum expectation value. In particular, the 
whole g2SC phase is replaced by the gluonic phase which 
is chromomagnetically stable.

We also compared the free energies of the 2SC/g2SC phase, 
the neutral LOFF state, and the gluonic phase. We found 
that the gluonic phase is energetically favored in the 
intermediate- and strong-coupling regimes. The encouraging 
results of this analysis should be further improved in the
future by 
(i) taking into account the most general ansatz for the 
gauge-field configuration in the gluonic phase,
(ii) by calculating the free energy in a self-consistent
manner. 
We already performed a preliminary investigation including the effect of 
the gluon field $D$, responsible for enforcing Gauss' law, 
and found that, for $\Delta_0 \alt 130$ MeV, the additional cost 
in the free energy is of order $0.01~{\rm MeV/fm}^3$, 
and thus negligible. For $\Delta_0 \agt 130$ MeV, however, 
the effect of the $D$ field could be an order of magnitude larger.

It is appropriate to mention that the instability related to the 
8th gluon was not studied in the present work. In this sense, the neutral 
LOFF state is appealing, because the Meissner mass of the 8th 
gluon is automatically zero in this state. It should also be 
mentioned that in the strongly coupled 
LOFF state \cite{Giannakis2004,Huang2005} 
the longitudinal Meissner mass squared of the 8th gluon 
is negative. Although this instability was not addressed 
in this work, it is unlikely, however, that the LOFF state is 
energetically favored in the strong-coupling regime.

\section*{Note added in proof}
It has recently been demonstrated 
that, in the three-flavor case, realistic crystal structures 
are more robust than a single plane-wave LOFF state \cite{Rajagopal2006}. 
In the two-flavor case, Bowers and Rajagopal \cite{Bowers2002} 
already indicated that a LOFF state with multiple plane waves 
would have a lower free energy than that with a single plane wave. 
The result shown in Fig. \ref{Figure3} would be altered by the inclusion of 
crystal structures with more plane waves. 
This is therefore an important project 
that needs to be addressed in future work.

\section*{Acknowledgments}

I.A.S. acknowledges discussions with V.A.~Miransky. 
This work was supported in part by the Virtual Institute of 
the Helmholtz Association under grant No. VH-VI-041, by the 
Gesellschaft f\"{u}r Schwerionenforschung (GSI), and by the 
Deutsche Forschungsgemeinschaft (DFG).

\end{document}